\documentclass[twocolumn,amsmath,amssymb,aps,pra,superscriptaddress,showpacs,tightenlines]{revtex4}
\usepackage{amsmath}
\usepackage{amsfonts}
\usepackage{graphicx}
\usepackage{epsfig}
\usepackage{color}
\usepackage{bm}
\usepackage[colorlinks,citecolor=blue]{hyperref}

\begin{document}

\title{Macroscopic quantum entanglement in modulated optomechanics}

\author{Mei Wang}
\author{Xin-You L\"{u}}
\email{xinyoulu@hust.edu.cn}
\affiliation{School of physics, Huazhong University of Science and Technology, Wuhan 430074, China}
\author{Ying-Dan Wang}
\affiliation{Institute of Theoretical Physics, Chinese Academy of Sciences, Beijing 100190, China}
\author{J. Q. You}
\affiliation{Quantum Physics and Quantum Information Division, Beijing Computational Science Research Center, Beijing 100193, China}
\author{Ying Wu}
\affiliation{School of physics, Huazhong University of Science and Technology, Wuhan 430074, China}
\date{\today}

\begin{abstract}
Quantum entanglement in mechanical systems is not only a key signature of macroscopic quantum effects, but has wide applications in quantum technologies.
Here we proposed an effective approach for creating strong steady-state entanglement between two directly coupled mechanical oscillators (or a mechanical oscillator and a microwave resonator) in a modulated optomechanical system.
The entanglement is achieved by combining the processes of a cavity cooling and the two-mode parametric interaction, which can surpass the bound on the maximal stationary entanglement from the two-mode parametric interaction.
In principle, our proposal allows one to cool the system from an initial thermal state to an entangled state with high purity by a monochromatic driving laser. Also, the obtained entangled state can be used to implement the continuous-variable teleportation with high fidelity. Moreover, our proposal is robust against the thermal fluctuations of the mechanical modes under the condition of strong optical pumping.
\end{abstract}

\maketitle

\section{Introduction}
Quantum entanglement~\cite{Horodecki2009} is a cornerstone of quantum
physics and has attracted wide interest in quantum technologies due to its potential applications in quantum information science~\cite{Braunstein2005,Jones2012} and quantum metrology~\cite{Giovannetti2011}.
To date quantum entanglement has been observed in various physical systems~\cite{Reid2009}, such as photonic~\cite{Pan2012}, atomic or molecular systems~\cite{Raimond2001}, and superconductor circuits~\cite{You2011},
ranging from microscopic systems to mesoscopic devices. It is desirable to realize macroscopic mechanical entanglement, because such entanglement might reveal the macroscopic quantum effects~\cite{Schwab2005} and then might possibly help us to clarify the boundary between classical and quantum worlds~\cite{Modi2012}.
Many methods have been proposed to prepare quantum entanglement in various mechanical systems~\cite{Eisert2004,Xue2007,Roncaglia2008,Huang2009,Jost2009,Ludwig2010,Cohen2013,Walter2013,xinyou2013,Szorkovszky2014,Johansson2014}.

Cavity optomechanics, exploring the interaction between the electromagnetic and mechanical modes, has progressed enormously in recent years~\cite{Review1,Review2,Review3}. It provides an alternative avenue to entangle two mechanical oscillators by exploiting the optomechanical radiation pressure~\cite{Mancini2002,Hartmann2008,Vacanti2008,Zhou2011,Joshi2012,Palomaki2013,Wang2015,Feng2015,Sete2014,Liao2014,Chen2014,Abdi2015}, using an optomechanical interferometer~\cite{Girvin2011,Zubairy2013,Zubairy,SeteJOSAB,Sete2015}, or the quantum interference in optomechanical interfaces~\cite{Tian2013}. More recently, the method of reservoir engineering is applied to the optomechanical systems (OMS), in order to prepare strongly entangled mechanical modes in the steady state~\cite{Wang2013,Tan2013,Woolley2014}. The main physical idea is to engineer the dissipation of the mechanical modes such that its steady state is the desired target state. Usually, more than two driving lasers are required to obtain the desired mechanical bath, which increases the difficulty of the practical implementation of the entanglement proposals.

Here we present a method to generate strong steady-state entanglement between two mechanical oscillators in an OMS via a two-mode parametric interaction and cavity cooling.
On one hand, the two-mode parametric interaction is induced by modulating an oscillator-oscillator coupling strength, which has been studied both theoretically~\cite{Tian2008,Lv2015} and experimentally~\cite{Westra2010,Westra2011,Okamoto2013}.
On the other hand, a red-detuned monochromatic laser is applied to the cavity, which generates strong linearized optomechanical coupling between the cavity and mechanical modes.
Interestingly, this monochromatic laser, when combined with the parametric process, could cool the mechanical system into a two-mode squeezed state from an initial thermal state.
We find that, near the optimal detuning points, the entanglement strength can go beyond the stationary entanglement limit, corresponding to a two-mode squeezing coupling (i.e., ${\rm ln}2$)~\cite{Wang2013}, even at high temperature. Compared with the previous studies, our proposal only requires one driving laser and is robust against the thermal fluctuations by increasing the driving power.
The obtained entangled state has high purity along with high entangled strength, which ensures the implementation of the standard continuous variable teleportation protocol with high fidelity.
Our proposal is general and can also be used to achieve the hybrid entanglement between a mechanical oscillator and a microwave resonator [see Fig.\,1(c)].
\begin{figure}[htb]
\centering\includegraphics[width=8.5cm]{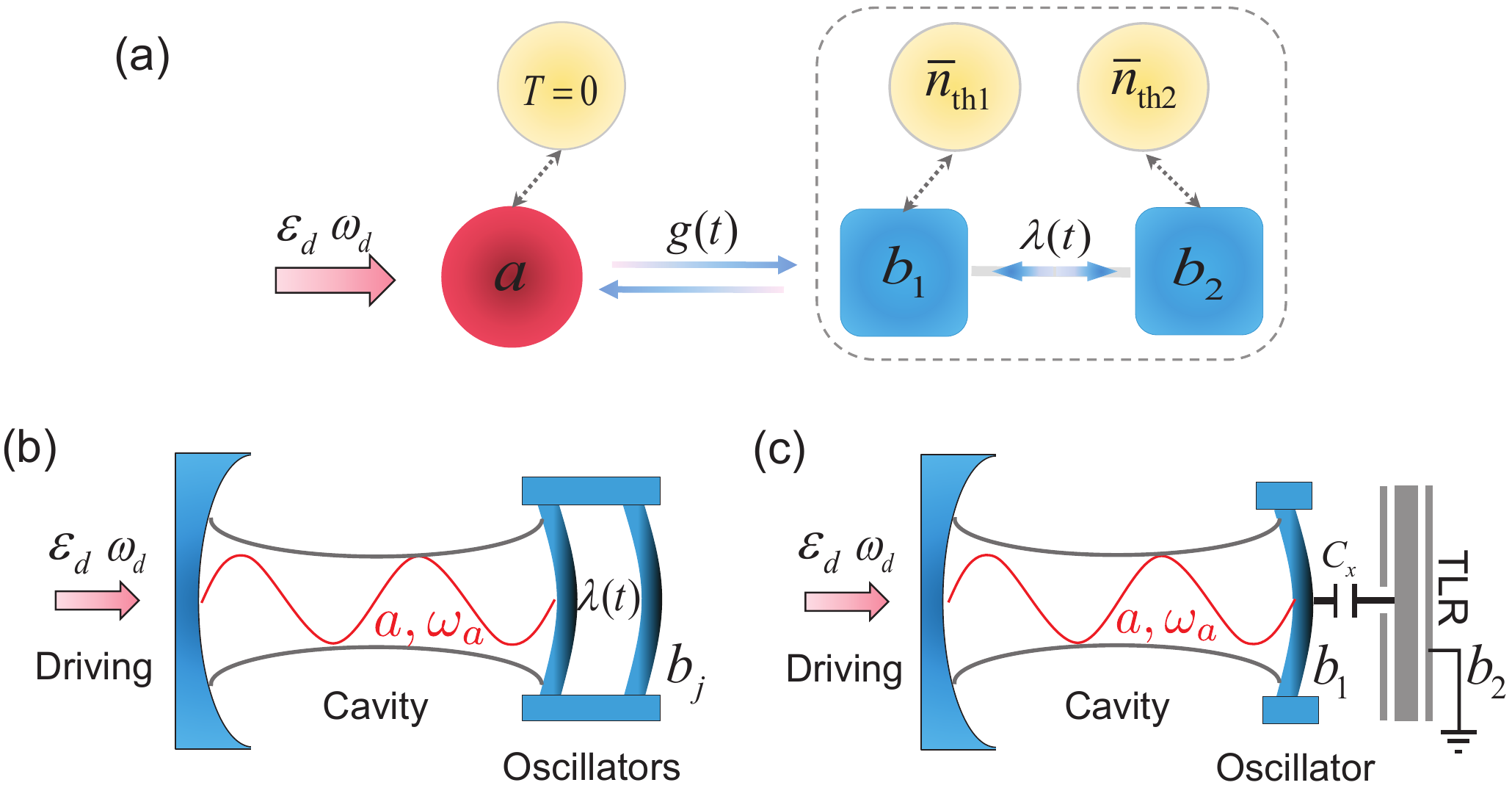}
\caption{(Color online) (a) A three mode boson system for preparing strong steady-state entanglement between modes $b_1$ and $b_2$.(b,c) The implementations of the above model in an OMS. An cavity mode $a$ (driven by a laser with frequency $\omega_d$ and amplitude $\varepsilon_d$) couples to (b) two interacted mechanical oscillators $b_j$ (j=1,2) or (c) a mechanical oscillator and a transmission line resonator (TLR) with a modulating coupling strength $\lambda(t)$. In panel (b), we assume that the cavity mode $a$ only couples to the first mechanical oscillator $b_1$ directly because $b_1$ is totally reflected. Here $g(t)$ indicates the modulated coupling strength between the optical mode $a$ and the mechanical mode $b_1$.}
\label{fig:1}
\end{figure}

This paper is organized as follows: In Sec. II, we introduce the general aspects of our proposal, which is implementary in an OMS with two directly coupled mechanical oscillators (or a mechanical oscillator coupled with a microwave resonator).
Then we derive its effective Hamiltonian under strong optical driving. In Sec. III, we study the steady-state entanglement between two mechanical oscillators (or a mechanical oscillator and a microwave resonator) and identify the optimal parameter regime for the maximum entanglement strength. In Sec. IV, we discuss the purity of the obtained entangled state and the teleportation fidelity when it is used as the entangled resource (``EPR channel'') in a continuous-variable teleportation protocol. Conclusions are given in Sec. VII.

\section{The Model}
We consider a three-mode boson system which couples each other via the modulated interactions with the forms of radiation-pressure and bilinearity, as shown in Fig.\,1(a). This general model is applicable to the OMS and can be used to prepare strong steady-state entanglement between two mechanical oscillators [see Fig.\,1(b)] or a mechanical oscillator and a microwave resonator [see Fig.\,1(c)].
Note that, modulating the optomechanical coupling strength can also be used to enhance the photon-phonon interactions~\cite{Liao2016}.

Without lossing the generality, our model can be considered as follows. Two mechanical oscillators (with frequencies $\omega_{\rm m1}$ and $\omega_{\rm m2}$) interact each other with a bilinear coupling $2\lambda_0 {\rm cos}(\omega_{\lambda}t)(b^{\dagger}_1+b_1)(b^{\dagger}_2+b_2)$. A cavity mode (with frequency $\omega_a$) is coupled to one of the mechanical modes with an optomechanical coupling
$2g_0{\rm cos}(\omega_{g}t)a^{\dagger}a(b^{\dagger}_{1}+b_1)$. Note that, this assumption is valid when the mechanical oscillator $b_1$ is totally reflected. Otherwise, the cavity mode $a$ also will couple to the second mechanical oscillator $b_2$ with the similar optomechanical coupling formation. However this will not change our result qualitatively, as shown in Sec. III. In this paper, we mainly discuss the model only including the interaction between the cavity mode $a$ and the first mechanical mode $b_1$. Here $\omega_{\lambda}$ and $\omega_{g}$ ($\lambda_0$ and $g_0$) are the modulating frequencies (amplitudes).
We use $a(a^{\dagger})$ and $b_j(b^{\dagger}_j)$ $(j=1,2)$ to denote the annihilation (creation) operators of the cavity mode and the mechanical modes, respectively. The cavity mode is driven by a monochromatic laser with frequency $\omega_d$ and amplitude $\varepsilon_d$ ($\varepsilon_{d}=\sqrt{2\kappa P/\hbar\omega _{d}}$ is related to the input power $P$ and the cavity decay rate $\kappa$).
\begin{figure}[htb]
\centerline{\includegraphics[width=8cm]{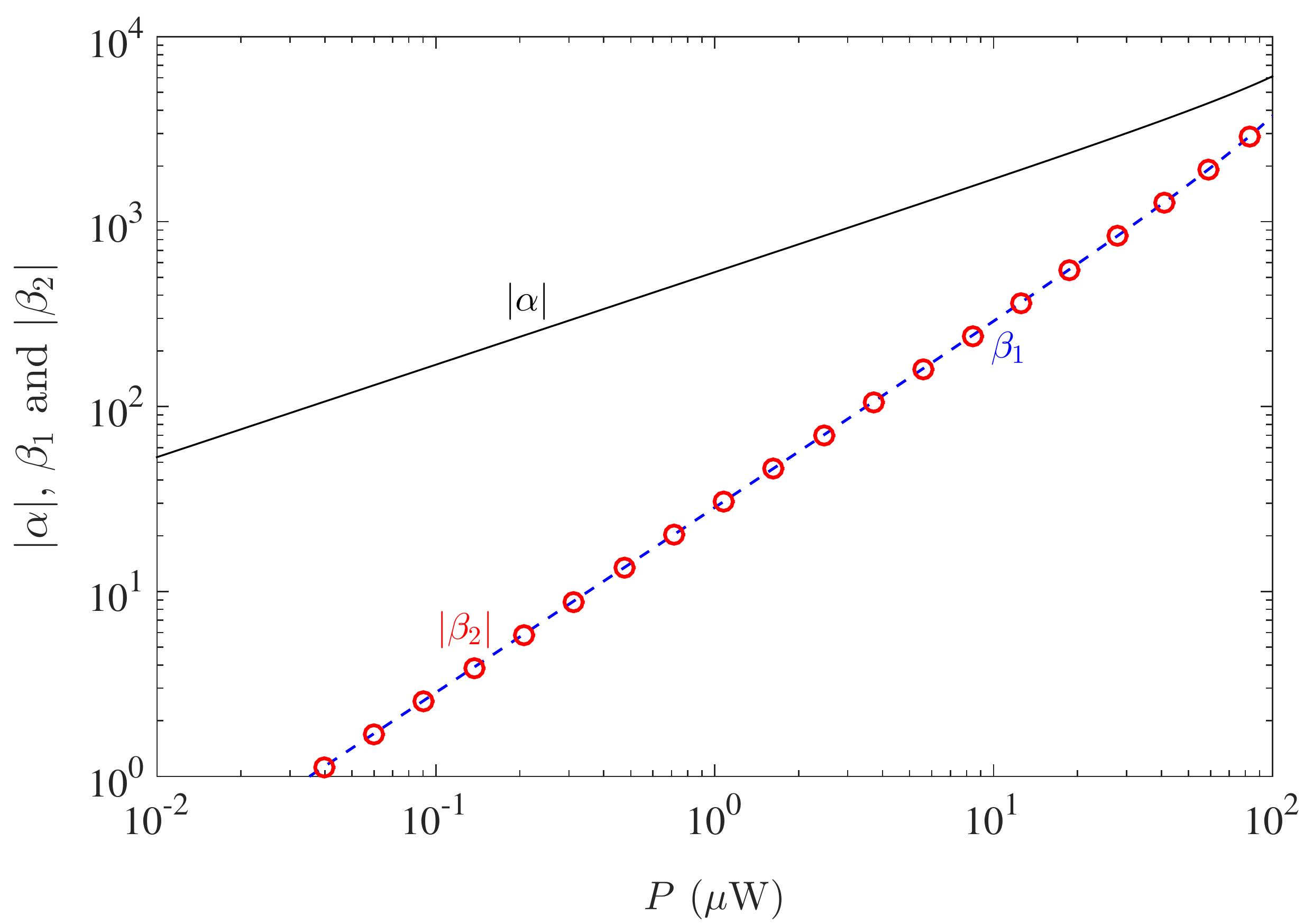}}
\caption{(Color online) The steady-state amplitudes $|\alpha|$, $\beta_{1}$ and $|\beta_{2}|$ versus driving power $P$. The parameters are $\omega_{d}=2\pi\times500$ THz, $\kappa=2\pi\times10^{5}$ Hz,  $g/\kappa=1\times10^{-4}$, $\lambda_0/\kappa=30$, $\Delta_1/\kappa=30.8$, $\Delta_2/\kappa=30.2$, $\Delta_a=\Omega'_1$ and $\gamma_{1}/\kappa=\gamma_{2}/\kappa=1\times10^{-5}$.}
\label{fig:2}
\end{figure}

\begin{figure}[htb]
\centerline{\includegraphics[width=8.6cm]{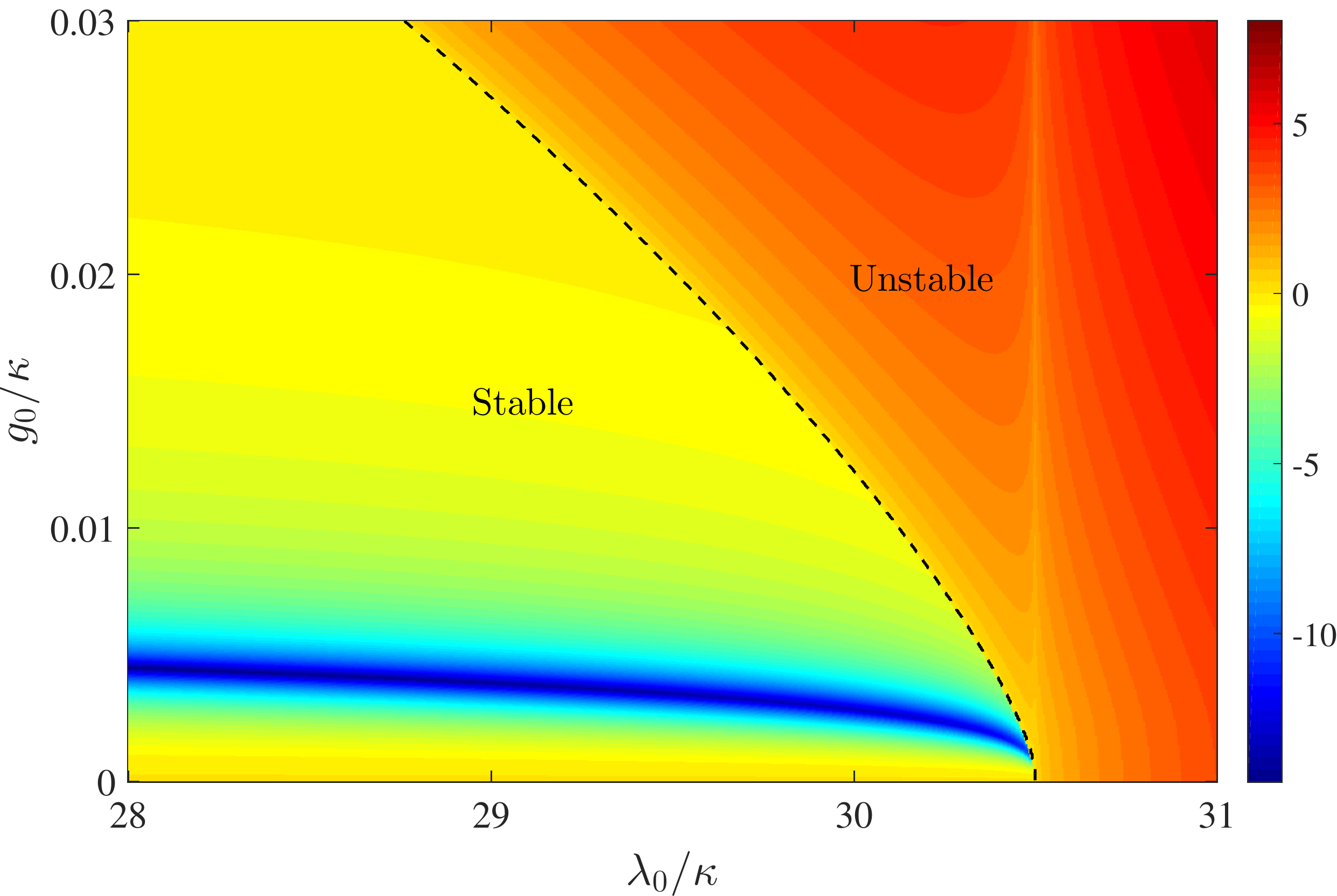}}
\caption{(Color online) The dependence of the standard of system stability on $g_0/\kappa$ and $\lambda_0/\kappa$. The black dashed line is the zero contour, separating the stable and the unstable regimes (In order to clearly see the distribution of the system stability, the values of the stable and unstable regimes are already diminished by four and six orders of magnitude, respectively.). Here $|\alpha|=10^2$, other parameters are same as that in Fig.\,2. And the optimal condition is the detuning $\Delta_a=\Omega'_1$.}
\label{fig:3}
\end{figure}

In a rotating frame with respective to the free Hamiltonian ${H}_{0}=\omega_{d}a^{\dagger}a+\omega_{g}b_{1}^{\dagger}b_{1}
+(\omega_{\lambda}-\omega_{g})b_{2}^{\dagger}b_{2}$, and considering the parameters condition $\omega_g$, $\omega_{\lambda}-\omega_g\gg g_0, \lambda_0$, the system Hamiltonian under the rotating-wave approximation (RWA) can be written as (we set $\hbar=1$)
\begin{eqnarray}
H_{\rm tot}&=&\delta_{a}a^{\dagger}a-g_0a^{\dagger}a(b_{1}^{\dagger}+b_{1})+\varepsilon_{d}(a^{\dagger}+a)+\Delta_{1}b_{1}^{\dagger}b_{1}
\nonumber\\&&
+\Delta_{2}b_{2}^{\dagger}b_{2}+\lambda_0(b_{1}^{\dagger}b_{2}^{\dagger}+b_{1}b_{2}),\label{e2}
\end{eqnarray}
where $\delta_{a}=\omega_{a}-\omega_{d}$, $\Delta_{1}=\omega_{\rm m1}-\omega_{g}$, and $\Delta_{2}=\omega_{\rm m2}+\omega_{g}-\omega_{\lambda}$ are the corresponding frequency detunings.

Including the dissipation caused by the system-bath coupling, the system dynamics is described by the Markovian master equation
\begin{eqnarray}
\dot{\rho}&=&-i[H_{\rm tot}, \rho]+\kappa \mathcal{D}[a]\rho
\nonumber\\
&&+\sum\limits_{j=1,2}\left[\gamma_{j}(\bar{n}_{\rm thj}+1)\mathcal{D}[b_{j}]\rho+\gamma_{j}\bar{n}_{\rm thj}\mathcal{D}[b_{j}^{\dagger}]\rho\right],\label{e3}
\end{eqnarray}
where $\mathcal{D}[o]\rho=o\rho o^{\dagger}-(o^{\dagger}o\rho+\rho o^{\dagger}o)/2$ ($o$ is a normal annihilation operator) is the standard Lindblad dissipative superoperator for the damping of the cavity (which is surrounded by a zero temperature environment) and mechanical modes. Here $\kappa$ and $\gamma_j$ are the cavity and the mechanical damping rates, respectively, and $\bar{n}_{\rm thj}$ is the thermal phonon occupation number.
\begin{figure}[tbh]
\centering
\includegraphics[width=8.6cm]{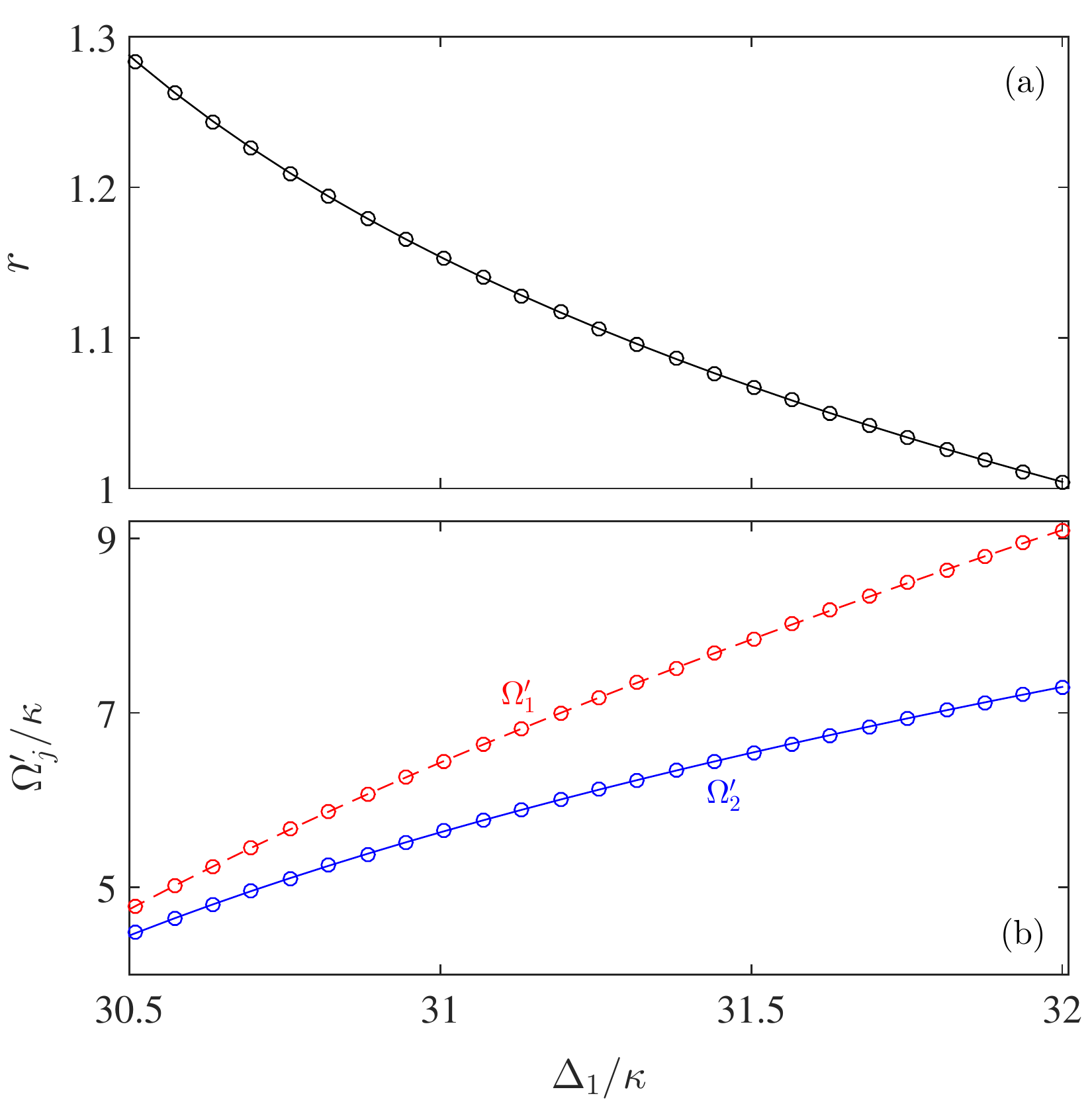}
\caption{(Color online) (a) The squeezing parameter $r$ and (b) transformed mechanical frequency $\Omega'_j/\kappa$ (j=1,2) (b) versus $\Delta_{1}/\kappa$. Parameters are the same as in Fig.\,2}
\label{fig:4}
\end{figure}

Strong red-detuned driving on the cavity generates large steady-state amplitudes in both the optical and mechanical modes. Following the standard linearization procedure, we can shift $a$ and $b_j$ with their steady-state mean values $\alpha$ and $\beta_j$, i.e., $a\rightarrow\alpha+a$, $b_{j}\rightarrow\beta_{j}+b_{j}$. The steady-state amplitudes $\alpha$ and $\beta_j$ can be derived by solving the following equations:
\begin{subequations}
\label{e3}
\begin{align}
&\Delta_{2}\beta_{2}+\lambda_0\beta_{1}^{*}=0,
\\
&\Delta_{1}\beta_{1}-g_0|\alpha|^{2}+\lambda\beta_{2}^{*}=0,
\\
&(\Delta_a-i\kappa/2)\alpha-2g_0\alpha {\rm Re}(\beta_{1})+\varepsilon_{d}=0,
\end{align}
\end{subequations}
where we have dropped the terms containing $\gamma_j$ because $\gamma_j\ll\kappa, \Delta_j, \lambda_0$. With strong optical driving on the cavity, the amplitudes $|\alpha|\gg1$, as shown in Fig.\,2. For example, with a driving power $P=3.5\times10^{-2}\,\mu$W, $\alpha\approx10^{2}$ could ensure the validity of our assumptions for linearization.

Then the nonlinear optomechanical coupling term $a^{\dagger}a(b_1^{\dagger}+b_1)$ can be ignored and the system dynamics is governed by
\begin{eqnarray}
\dot{\rho}&=&-i[H_{\rm li}, \rho]+\kappa \mathcal{D}[a]\rho
\nonumber\\
&&+\sum\limits_{j=1,2}\left[\gamma_{j}(\bar{n}_{\rm thj}+1)\mathcal{D}[b_{j}]\rho+\gamma_{j}\bar{n}_{\rm thj}\mathcal{D}[b_{j}^{\dagger}]\rho\right],\label{e4.5}
\end{eqnarray}
with the linearized Hamiltonian
\begin{eqnarray}
H_{\rm li}&=&\Delta_{a}a^{\dagger}a+\sum\limits_{j=1,2}\Delta_{j}b_{j}^{\dagger}b_{j}-G(a^{\dagger}+a)(b_{1}^{\dagger}+b_{1})
\nonumber\\&&
+\lambda_0(b_{1}^{\dagger}b_{2}^{\dagger}+b_{1}b_{2}).\label{e5}
\end{eqnarray}
Here $\Delta_{a}=\delta_{a}-2{\rm Re}(\beta_{1})g_0$ and $G=g_0|\alpha|$ are the shifted detuning and the linearized optomechanical coupling, respectively. The third term of $H_{\rm li}$ describes a linearized optomechanical interaction and provides the cooling process. combining with the phonon-phonon parametric coupling (the last term) can cool the mechanical modes $b_1$ and $b_2$ into a two-mode squeezed state in the steady state. Compared with the normal OMS, here the relative low driving power can induce large optical and mechanical amplitudes $|\alpha|$ and $|\beta_j|$ (see Fig.\,2) because of the two-mode parametric amplification term, i.e., the last term in Eq.\,(1).

 In other words, parametric-amplification process can also induce instability. In Fig.\,3, we show the numerical stability condition for the system with $|\alpha|=10^2$. In our system, the weak cavity-resonator coupling $g_0/\kappa=1\times10^{-4}$ and the resonator-resonator coupling $\lambda_0/\kappa=30$ are used in the stable regime. Meanwhile, our parameter regimes are well separated from the bistability threshold for a Duffing oscillator.

\section{Mechanical entanglement}
We apply the two-mode squeezing transformation $S(r)={\rm exp}[r(b_{1}b_{2}-b_{1}^{\dagger}b_{2}^{\dagger})]$, with squeezing parameter
\begin{eqnarray}
r=\frac{1}{4}\ln\left(\frac{\Delta_{1}+\Delta_{2}+2\lambda_0}{\Delta_{1}+\Delta_{2}-2\lambda_0}\right),\label{e5.5}
\end{eqnarray}
to the Hamiltonian $H_{\rm li}$. Using the squeezing transformation $S^{\dagger}(r)aS(r)=a$ and
\begin{subequations}
\label{e7}
\begin{align}
S^{\dagger}(r)b_{1}S(r)=b_{1}{\rm cosh}(r)-b_{2}^{\dagger}{\rm sinh}(r),\label{e7a}
\\
S^{\dagger}(r)b_{2}S(r)=b_{2}{\rm cosh}(r)-b_{1}^{\dagger}{\rm sinh}(r),\label{e7b}
\end{align}
\end{subequations}
the Hamiltonian is hence transformed to $H'_{\rm li}=S^{\dagger}(r)H_{\rm li}S(r)$, with
\begin{eqnarray}
\label{e8}
H'_{\rm li}&=&\Delta_{a}a^{\dagger}a+\Omega'_{1}b_{1}^{\dagger}b_{1}+\Omega'_{2}b_{2}^{\dagger}b_{2}
-G'_{1}(a^{\dagger}+a)(b_{1}^{\dagger}+b_{1})
\nonumber\\&&
+G'_{2}(a^{\dagger}+a)(b_{2}^{\dagger}+b_{2}).\label{e8}
\end{eqnarray}
Here $G'_{1}=G{\rm cosh}(r)$ and $G'_{2}=G{\rm sinh}(r)$ are the transformed optomechanical couplings. The transformed mechanical frequencies are
\begin{subequations}
\label{e9}
\begin{eqnarray}
\!\!\!\!\Omega'_{1}\!\!=\!\!\Delta_{1}{\rm cosh}^{2}(r)\!+\!\Delta_{2}{\rm sinh}^{2}(r)\!-\!2\lambda_0{\rm cosh}(r){\rm sinh}(r)\!,
\\
\!\!\!\!\Omega'_{2}\!\!=\!\!\Delta_{1}{\rm sinh}^{2}(r)\!+\!\Delta_{2}{\rm cosh}^{2}(r)\!-\!2\lambda_0{\rm cosh}(r){\rm sinh}(r)\!,
\end{eqnarray}
\end{subequations}
which are decided by the frequency detunings $\Delta_{j}$ and the coupling strength $\lambda_0$. As shown in Fig.\,4, large squeezing parameter $r$ can be obtained in our proposal, which ensures high entanglement well beyond the limit ${\rm ln}2$. At the same time, the relatively large values of $\Omega'_1$ and $\Omega'_2$ obtained here effectively suppress the quantum backaction noise from the optical cavity to the mechanical modes during the cooling process.

Considering the system-bath coupling, we apply the two-mode squeezing transformation $S(r)$ to the master equation (4) and define the transformed density matrix $\rho'=S^{\dagger}(r)\rho S(r)$. Under the condition $\Delta_a$, $\Omega'_j\gg G'_j,\gamma_j(\bar{n}_{\rm th}+1)$, the counter-rotating terms in dissipations, $\mathcal{G}[b_{12}]\rho'$ and $\mathcal{G}[b^{\dagger}_{12}]\rho'$, are fast oscillating with factors $\sim e^{\pm i(\Omega'_1+\Omega'_2)t}$ and can be neglected safely. Here we have used the definition $\mathcal{G}[b_{12}]\rho=b_1\rho b_2-(b_1b_2\rho+\rho b_1b_2)/2$.
Then under the RWA, the transformed master equation for $\rho'$ has the same form as Eq.\,(4), with $H_{\rm li}$ replaced by $H'_{\rm li}$; $\gamma_{j}$ and $\bar{n}_{\rm thj}$ by
\begin{subequations}
\label{e9.5}
\begin{eqnarray}
\!\!\gamma_1'&\!\!=\!\!&\gamma_1{\rm cosh}^{2}(r)-\gamma_2{\rm sinh}^{2}(r),
\\
\!\!\gamma_2'&\!\!=\!\!&\gamma_2{\rm cosh}^{2}(r)-\gamma_1{\rm sinh}^{2}(r),
\\
\!\!\bar{n}^{'}_{\rm th1}&\!\!=\!\!&\frac{\gamma_1\bar{n}_{\rm th1}{\rm cosh}^2(r)+\gamma_2(\bar{n}_{\rm th2}+1){\rm sinh}^2(r)}{\gamma_1{\rm cosh}^{2}(r)-\gamma_2{\rm sinh}^{2}(r)},
\\
\!\!\bar{n}^{'}_{\rm th2}&\!\!=\!\!&\frac{\gamma_2\bar{n}_{\rm th2}{\rm cosh}^2(r)+\gamma_1(\bar{n}_{\rm th1}+1){\rm sinh}^2(r)}{\gamma_2{\rm cosh}^{2}(r)-\gamma_1{\rm sinh}^{2}(r)}.
\end{eqnarray}
\end{subequations}
This transformed master equation for $\rho'$ describes a standard cavity cooling process for the two mechanical oscillators, which are decoupled in the transformed representation (see the Hamiltonian $H'_{\rm li}$).

Qualitatively, the proposed entanglement scheme can be better understood in the cooling regime $\Omega'_j\gg\kappa\gg G'_{j}$ which yields a simple analytical solution.
A cooling equation for the mechanical modes can be derived from the master equation in the transformed basis by adiabatically eliminating the cavity
mode \cite{Kippenberg2007,Marquardt2007,Vitali2008}. By defining $\rho'_m={\rm Tr}_a[\rho']$ as the reduced density matrix of the mechanical modes, the cooling master equation is
\begin{eqnarray}
\dot{\rho}'_{m}&\approx&-i[H'_{m},\rho'_{m}]+\sum\limits_{j=1,2}\left\{[\gamma'_{j}(\bar{n}'_{\rm thj}+1)+\Gamma^{-}_{j}]\mathcal{D}[b_{j}]\rho'_{m}\nonumber\right.
\\
&&\left.+(\gamma'_{j}\bar{n}'_{\rm thj}+\Gamma^{+}_{j})\mathcal{D}[b_{j}^{\dagger}]\rho'_{m}\right\},
\end{eqnarray}
where $H'_{m}=\tilde{\Omega}'_{1}b_{1}^{\dagger}b_{1}+\tilde{\Omega}'_{2}b_{2}^{\dagger}b_{2}$, with the effective mechanical frequency $\tilde{\Omega}'_{j}$ $(j=1,2)$. Under the parameter condition considered here, $\tilde{\Omega}'_{j}\approx\Omega'_j$ and the RWA has been applied during the derivation of the above equation.
The cavity induced cooling and heating rates $\Gamma^{\mp}_{j}$ are given by
\begin{align}
\label{e12}
\Gamma^{\mp}_{j}=\frac{\kappa (G'_{j})^2}{\kappa^{2}/4+(\tilde{\Delta}_{a}\mp\Omega'_{j})^{2}}.
\end{align}
The steady state of Eq.\,(11) is a two-mode thermal state with average phonon number
\begin{eqnarray}
\bar{n}'_{\rm effj}=\frac{\gamma'_{j}\bar{n}'_{\rm thj}+\Gamma^{+}_{j}}{\gamma'_{j}+\Gamma_{j}},\label{e15}
\end{eqnarray}
where $\Gamma_j=\Gamma^{-}_j-\Gamma^{+}_j$ is the net cooling rate. It shows that the minimal $\bar{n}'_{\rm effj}$ can be obtained by the optimal detuning $\Delta_a=\Omega'_j$, which corresponds to $\Gamma^{-}_j=4(G'_j)^2/\kappa$ and $\Gamma^{+}_j\approx\kappa[G'_j/(2\Omega'_j)]^2$. Hence the mechanical mode in the original basis is in a two-mode squeezed thermal state. The entanglement degree depends on the squeezed parameter $r$ and the cooling rate $\Gamma$, which are decided by the driving laser and the above modulated coupling [see Eq.\,(6)]. For an ideal case, ignoring the quantum backaction noise, the mechanical oscillators can be cooled into a two-mode vacuum state in the transformed representation $|00\rangle$. In the original representation, this is a two-mode squeezed state $S(r)|00\rangle$, whose logarithmic negativity is $E_N=2r$. Then, the entangled degree is enhanced by adjusting the squeezed parameter $r$ via the tunable system parameters $\Delta_j$ and $\lambda_0$, as shown in Eq.\,(6) and Fig.\,4(a).
\begin{figure}[tbh]
\centering
\includegraphics[width=8cm]{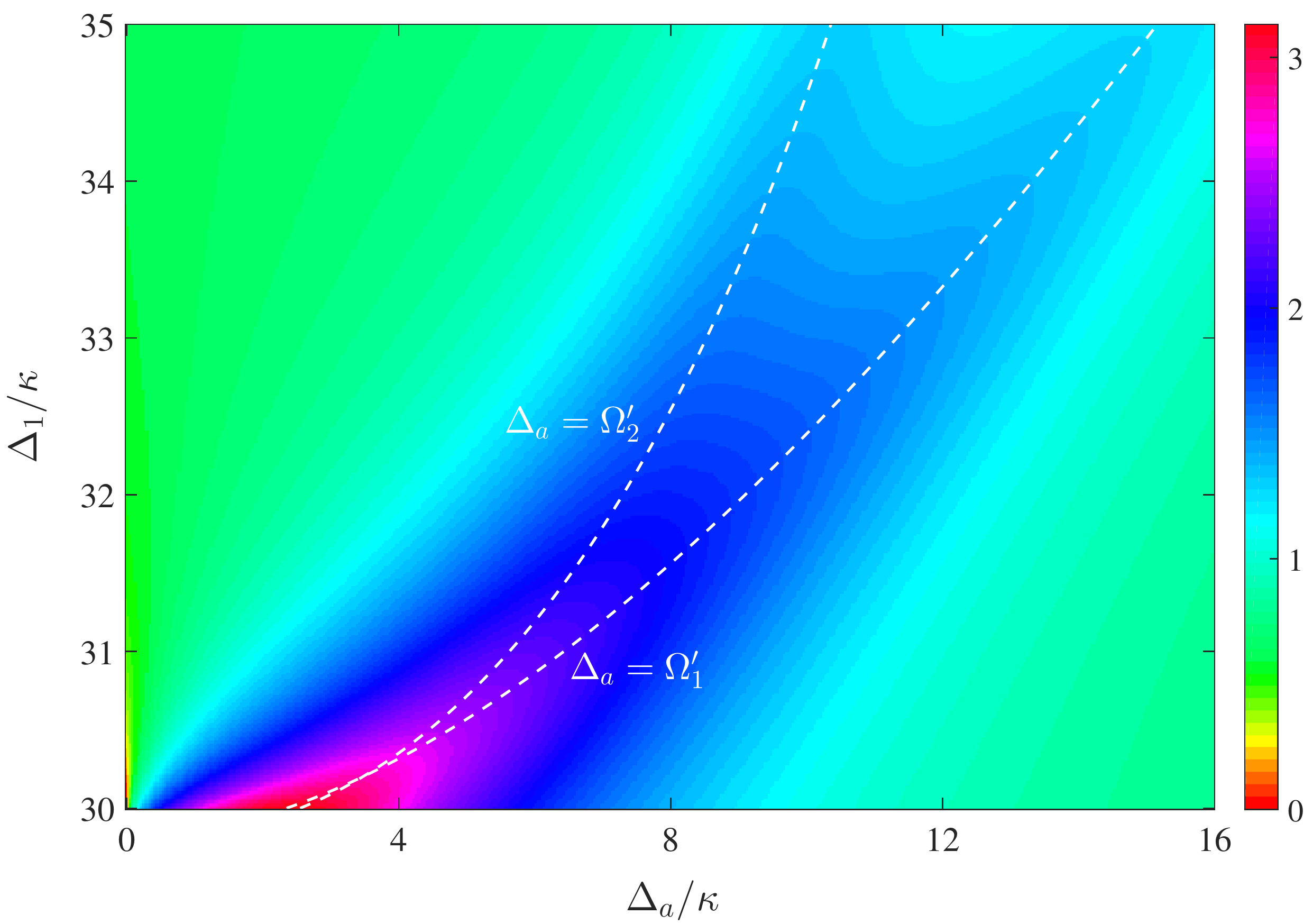}
\caption{(Color online) The entanglement degree $E_{N}$ of two mechanical oscillators versus $\Delta_{1}$ and $\Delta_{a}$. Here, $|\alpha|=10^2$, $P=3.5\times10^{-2}\mu$W, other parameters are same as that in Fig.\,2. The dashed lines corresponds to entanglement at the optimal detunings, i.e., $\Delta_{a}=\Omega'_{j}$ $(j=1,2)$.}
\label{fig:5}
\end{figure}

\begin{figure}[tbh]
\centerline{\includegraphics[width=8.5cm]{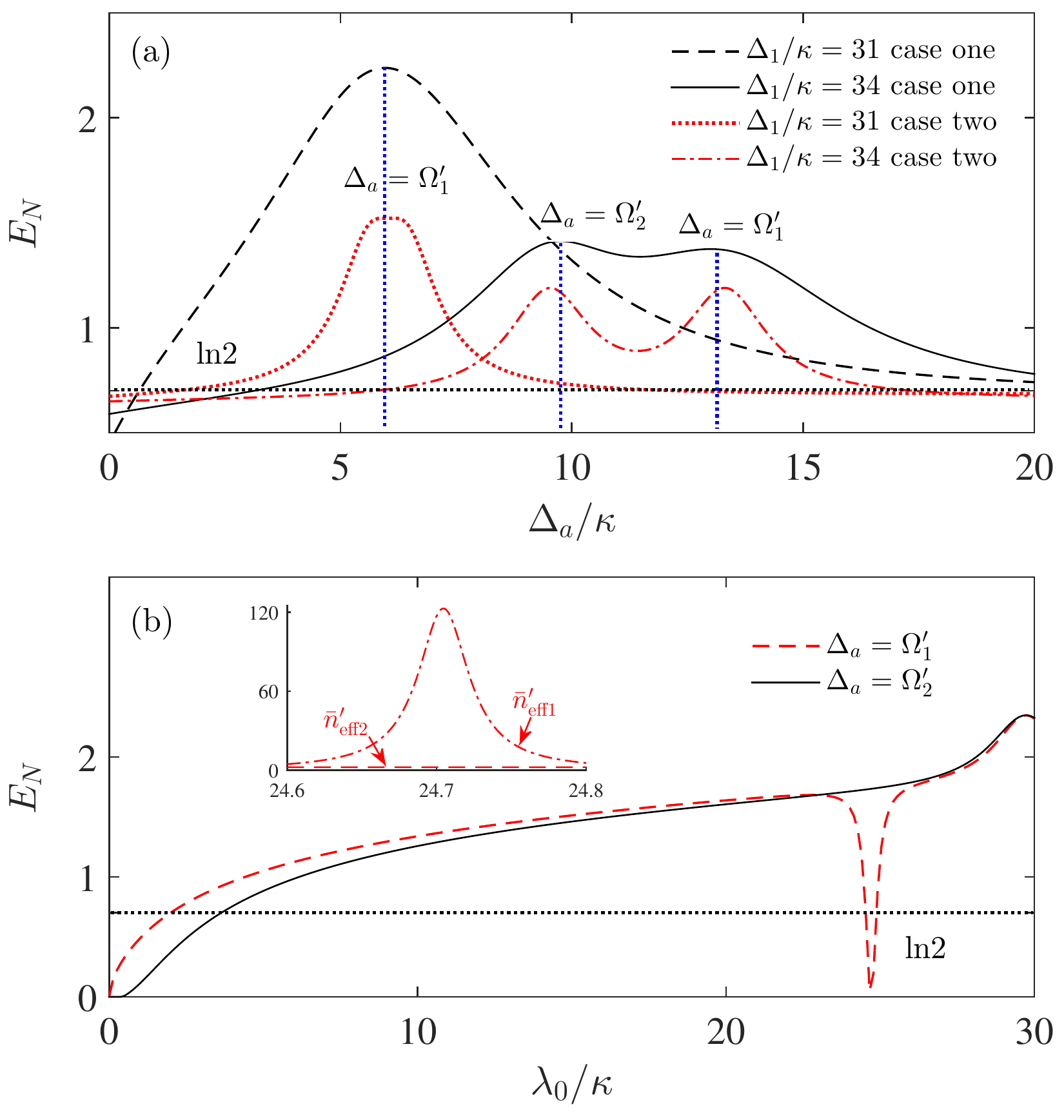}}
\caption{(Color online) The entanglement degree $E_N$ versus (a) the cavity detuning $\Delta_a$, and (b) the coupling strength $\lambda_0$. In (a), the black dashed and the black solid curves correspond to the model only including the coupling between the cavity mode and the first mechanical mode (case one). The red dotted and red dot-dashed curves correspond to the model including the interaction between the cavity mode and two mechanical modes (case two). The optimal detuning $\Delta_a=\Omega'_1$ or $\Omega'_2$ is chosen in (b). The inserts in (b) indicate the average phonon number $\bar{n}'_{\rm effj}$ (j=1,2) corresponding to $\Delta_a=\Omega'_1$ (red lines). The parameters are the same as in Fig.\,2 except for $\bar{n}_{\rm th}=0$ and $\Delta_{1}+\Delta_{2}-2\lambda_0=1$.}
\label{fig:6}
\end{figure}

\begin{figure}[tbh]
\centerline{\includegraphics[width=8.5cm]{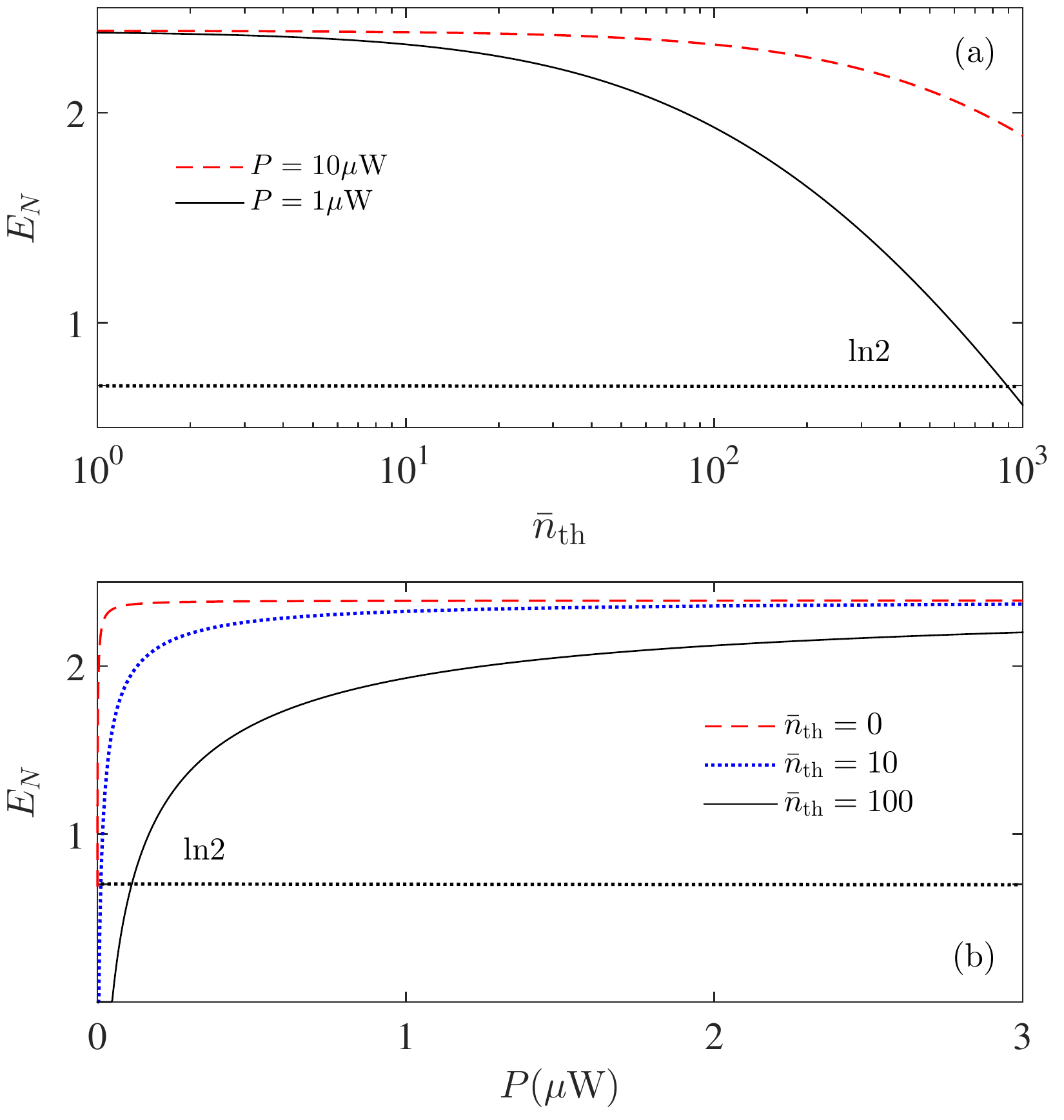}}
\caption{(Color online) The logarithmic negativity $E_N$ versus (a) the thermal phonon number $\bar{n}_{\rm th}$, and (b) the driving power $P$ when the optimal detuning is chosen, i.e., $\Delta_a=\Omega'_1$.
The system parameters are the same as in Fig.\,2.}
\label{fig:7}
\end{figure}

To support the qualitative discussion, we now calculate the logarithmic negativity $E_N$ quantifying the degree of mechanical entanglement based on the shifted master equation (4).
Here the logarithmic negativity $E_N$ is defined as $E_{N}={\rm max}[0, -\ln(2\eta^{-})]$~\cite{Vidal2002}, which is decided by the covariance matrix ${\rm \textbf{V}}$ according to
\begin{eqnarray}
\eta^{-}=\frac{1}{\sqrt2}\sqrt{\Sigma-\sqrt{\Sigma-4{\rm det} {\rm \textbf{V}}}}
\end{eqnarray}
with
\begin{eqnarray}
\Sigma={\rm det} {\rm \textbf{B}}+{\rm det} {\rm \textbf{B}'}-2{\rm det} {\rm \textbf{C}}.
\end{eqnarray}
Here ${\rm \textbf{V}}$ is a 4$\times$4 covariance matrix of the two mechanical modes, defined as ${\rm \textbf{V}}_{jk}=\frac{1}{2}\langle\Delta\xi_{j}\Delta\xi_{k}+\Delta\xi_{k}\Delta\xi_{j}\rangle$ with $\Delta\xi_{j}=\xi_{j}-\langle\xi_{j}\rangle$, $\vec{\xi}=\{x_{1}, p_{1}, x_{2}, p_{2}\}$, where $x_{j}=(b_{j}+b_{j}^{\dagger})/\sqrt2$, and $p_{j}=-i(b_{j}-b_{j}^{\dagger})/\sqrt2$. Here ${\rm \textbf{B}}$,
${\rm \textbf{B}'}$ and ${\rm \textbf{C}}$ are 2$\times$2 matrices in
\begin{equation}
{\rm \textbf{V}}=
\left(
  \begin{array}{cc}
    {\rm \textbf{B}} & {\rm \textbf{C}} \\
    {\rm \textbf{C}}^{T} & {\rm \textbf{B}'} \\
  \end{array}
\right).
\end{equation}

The numerical results in Figs.\,5 and \,6(a) clearly demonstrate that, at a given squeezing parameter $r$, entanglement is the strongest at the optimal detuning $\Delta_a=\Omega'_j$ (j=1,2). This corresponds to the best cooling
for the mechanical modes in the transformed representation. This numerical result is consistent with our qualitative discussion obtained in the cooling limit. In Fig.\,6(a), we also present the result when the cavity mode $a$ couples to two mechanical modes simultaneously (case two). It is shown that there is not much difference between our model (i.e., case one) and the case two, except for the entanglement degree decrease a little in case two. Physically, in our proposal the entanglement strength is decided by the optical cooling capacity in the squeezed representation, i.e., the net cooling rate $\Gamma_j$. With the similar calculations for deriving Eq.\,(12), one can obtain that, in case two, $\Gamma_j$ (being proportional to $e^{-2r}$) is smaller than that in case one ($\Gamma_j$ being proportional to $e^{2r}$) in the same parameter regime. This leads to the result that the entanglement degree becomes smaller in case two comparing with that in case one. In Fig.\,6(b), we plot $E_N$ as a function of the mechanical coupling strength $\lambda_0$ under the conditions of $\Delta_a=\Omega'_j$ (at zero temperature $\bar{n}_{\rm th}=0$, $\bar{n}_{\rm th1}=\bar{n}_{\rm th2}=\bar{n}_{\rm th}$). Our result shows that, as the mechanical coupling strength reaches a threshold value, the entanglement degree can exceed the steady-state entanglement limit, ${\rm ln}2$, from the parametric interaction. Moreover, there is minimal value of $E_N$, corresponding a maximum value of $\bar{n}'_{\rm effj}$, along with increasing $\lambda_0$. This originally comes from the competition between the cooling rate $\Gamma^-_{j}$ and the heating rate $\Gamma^+_{j}$, when the optimal detuning is chosen. Specifically, when increasing $\lambda_0$, the competition between the increasing optomechanical coupling $G'_j$ and the decreasing transformed mechanical frequency $\Omega'_{j}$ leads to a minimal net cooling rate $\Gamma$. In Fig.\,7, we present the influences of the mechanical thermal noise $\bar{n}_{\rm th}$ and the driving power $P$ on the entanglement degree. It shows that, even at a high temperature with $\bar{n}_{\rm th}=1\times10^2$, strong steady-state entanglement can still be reached by increasing the driving power. In our proposal, the strong optical driving induces strong cooling efficiency of the cavity, which suppresses the influence of the mechanical thermal bath on the entanglement degree.

Here we would like to emphasize the physical mechanism of our proposal clearly, which is quite different from the previous studies. In our proposal, the mechanical modes can be cooled into a two-mode thermal state in the squeezing representation, which corresponds to a two-mode squeezed thermal state in the original basis. Then given a fixed squeezing strength (i.e., squeezing parameter $r$), the better of the cooling efficiency the higher entanglement degree can we get. It is shown from Eq.\,(12) that the optimal detuning $\Delta_a=\Omega'_j$ corresponds to the maximal cooling rate. This ultimately leads to the result that the maximal entanglement between the mechanical oscillators is at $\Delta_a=\Omega'_j$ in our proposal.

\begin{figure}[tbh]
\centerline{\includegraphics[width=8.5cm]{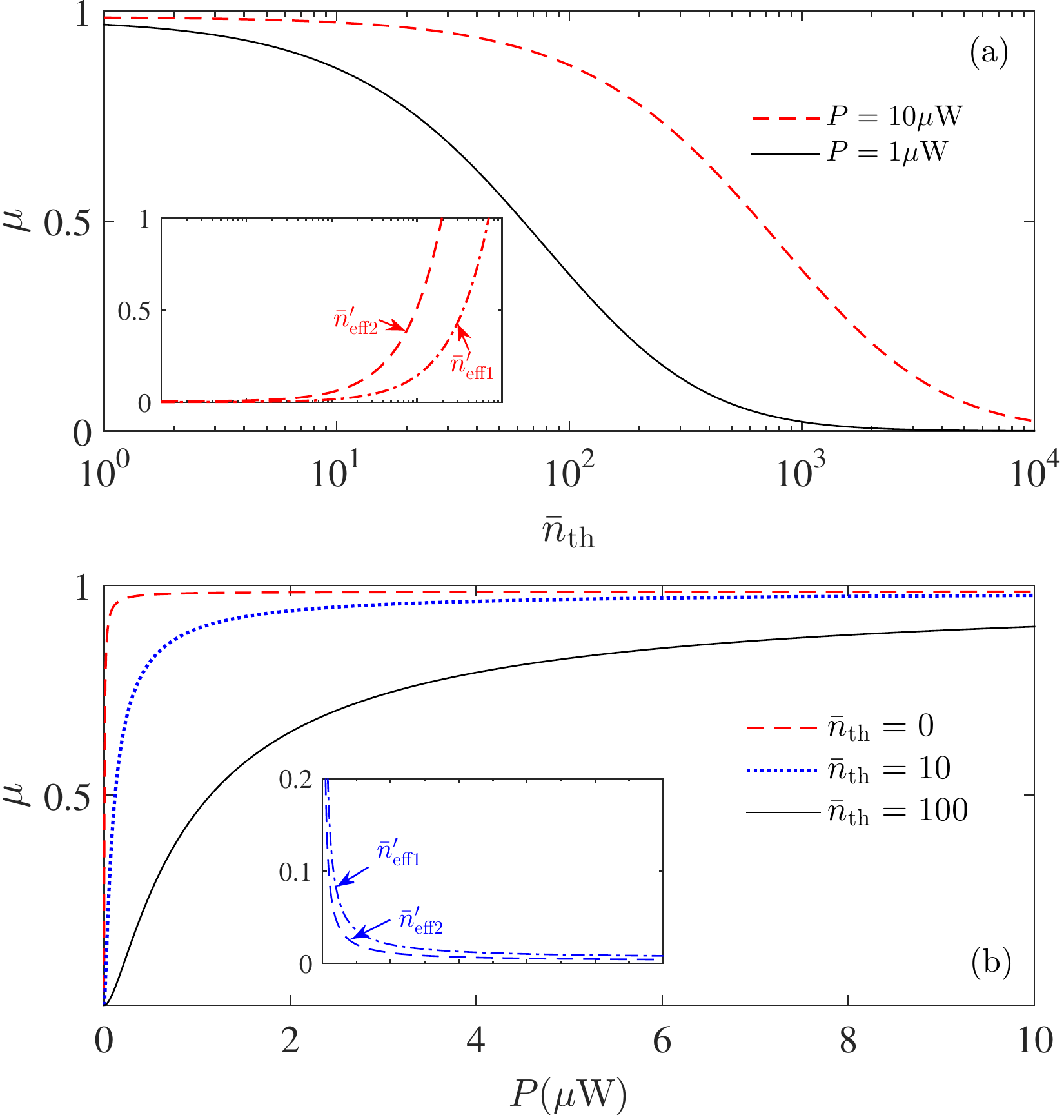}}
\caption{(Color online) The purity of steady-state $\mu$ versus (a) the thermal occupation $\bar{n}_{\rm th}$ and (b) the driving power $P$.
The inserts indicate the average phonon number $\bar{n}'_{\rm effj}$ (j=1,2) corresponding to (a) $P=10\mu$W and (b) $\bar{n}_{\rm th}=10$. Parameters are the same as in Fig.\,7.}
\label{fig:8}
\end{figure}

\begin{figure}[tbh]
\centerline{\includegraphics[width=8.5cm]{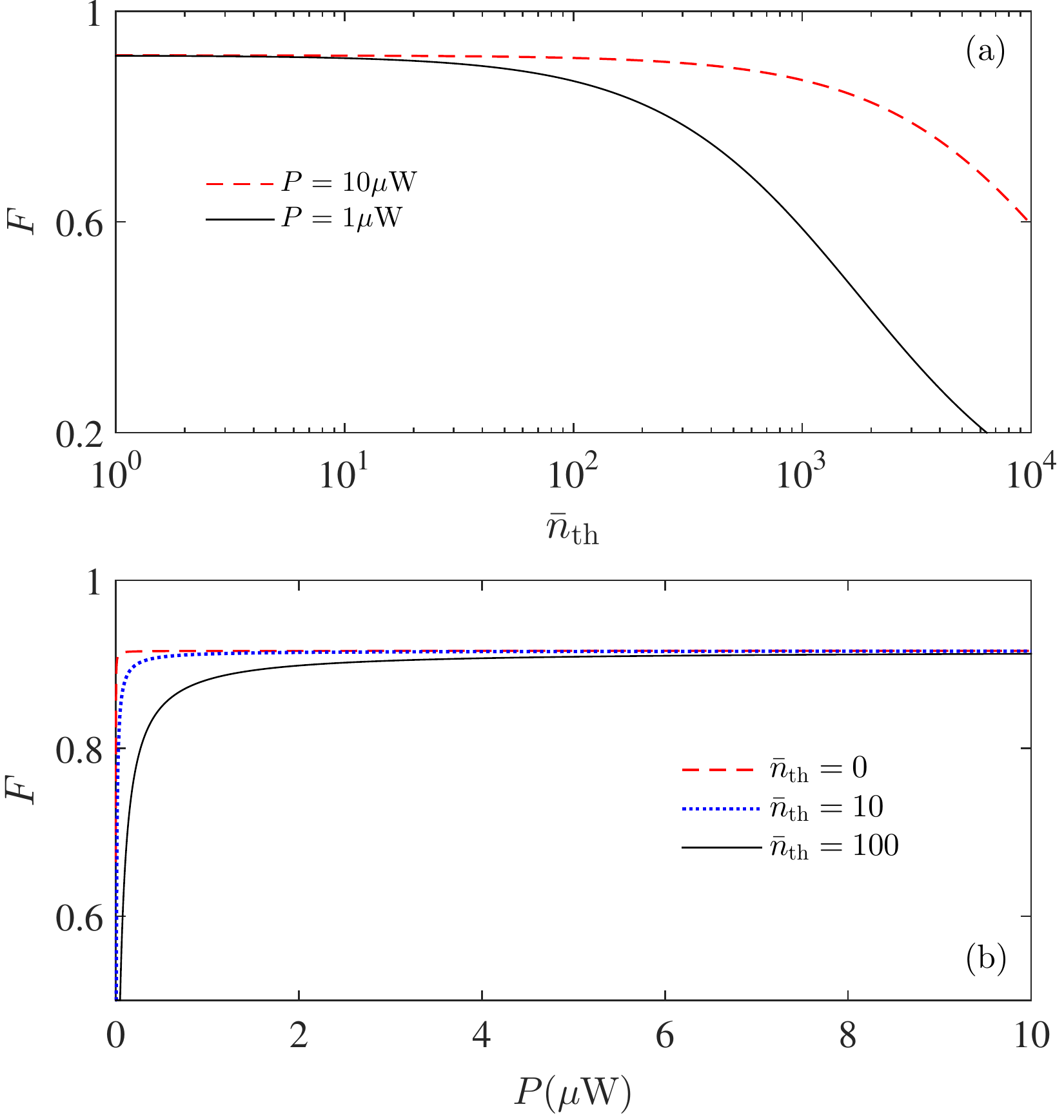}}
\caption{(Color online) (a) The teleportation fidelity $F$ of the two-mode squeezed state versus (a) the thermal phonon number $\bar{n}_{\rm th}$, and (b) the driving power $P$. The system parameters are the same as in Fig.\,2.}
\label{fig:9}
\end{figure}

\section{Discussions}
For an efficient application of the entangled state to the modern quantum technologies, the purity of the steady state is the foundation. For example, in a continuous-variable teleportation protocol, the entangled state generated here may be regarded as the entangled resource (¡°EPR channel¡±), and the high purity corresponds to a high fidelity~\cite{Braunstein1998}.

In the sections above, we can get highly entangled steady state; however, it does not make sure that the steady state is also highly pure.
In the cooling limit, the mechanical system in
the original basis is a two-mode squeezed thermal state in the steady state. Then let us define the purity $\mu$ of the steady state as
\begin{eqnarray}
\mu={\rm tr}(\rho'^{2}_{m}),
\end{eqnarray}
where $\rho'_{m}$ is the reduced density matrix of the two mechanical modes after the adiabatic elimination. With the covariance matrix ${\rm \textbf{V}}$ defined above, the purity can also be simplified as~\cite{Woolley2014}
\begin{eqnarray}
\mu&\!\!=\!\!&1/(4\sqrt{\rm det\textbf{V}})
\nonumber\\
&\!\!=\!\!&\frac{1}{(1+2\bar{n}'_{\rm eff1})(1+2\bar{n}'_{\rm eff2})}.
\end{eqnarray}
Equation (18) clearly shows that the purity is inversely proportional to the steady-state phonon numbers of the transformed system, which are determined by the cooling efficiency. Through the cooling master equation, we numerically calculate the purity of the obtained entangled state in Fig.\,8. It shows that highly purity can be obtained in the considered parameter regime corresponding to the highly entanglement degree. The purity is also robust against the initial mechanical noise featured by the thermal occupations $\bar{n}_{\rm th}$. Physically, in our proposal, a large squeezing parameter $r$ allows us to obtain the approximately equal cooling rates for two mechanical modes (see the definition of $\Gamma_j$). Then, as shown in Eq.\,(13) and the inserts in Fig.\,8, two mechanical oscillators can be cooled down simultaneously in the transformed system under the condition of strong optical driving.
This ultimately leads to the result that the steady state of system will has a high purity after a strong cavity cooling process.

Generally, a continuous-variable teleportation protocol can be implemented with high fidelity when a highly pure entangle-state is served as the ¡°EPR channel¡±.
Here we actually obtain a two-mode squeezed thermal state in the steady state and then the corresponding teleportation fidelity can be written as~\cite{Fiurasek2002}
\begin{eqnarray}
F=\frac{1}{e^{-2r}(1+\bar{n}'_{\rm eff1}+\bar{n}'_{\rm eff2}+e^{2r})}.
\end{eqnarray}
Clearly, the larger $\bar{n}'_{\rm effj}$ leads to a lower purity [refer to Eq.(18)] and a lower teleportation fidelity.
The best cooling in the transformed system occurs at the optimal detuning $\Delta_a=\Omega'_{j}$. Hence, Eq.\,(19) shows that, at a given amount of entanglement, the teleportation fidelity is highest at the optimal detuning.

Under the best cooling conditions, we present the influences of the thermal occupation $\bar{n}_{\rm th}$ and the driving power $P$ on the teleportation fidelity in Fig.\,9.
Consisting with our qualitative discussion, in our proposal, high teleportation fidelity could be obtained when using a strong driving laser to cool the mechanical modes simultaneously. The effective thermal phonon number $\bar{n}'_{\rm effj}$ is close to be zero under the optimal parameters regime, as shown in Fig.\,8. Naturally, the fidelity can reach $0.92$ easily. In addition, our results also show that, even at a high temperature $\bar{n}_{\rm th}=1\times10^{2}$, the high teleportation fidelity can still be achieved by increasing the driving power.

\section{Conclusion}
We have provided a method to generate strong steady-state entanglement between two mechanical oscillators (or a mechanical oscillator and a microwave resonator) that is robust against the thermal fluctuations. Our approach utilizes a modulated phonon-phonon (or phonon-photon) interaction and a strong driving on the cavity mode in an OMS. The entanglement is a consequence of the joint effect of the two-mode parametric interaction and cavity cooling. We have showed that strong entanglement can be achieved at the optimal detuning where the cavity detuning is in resonance with one of the transformed mechanical frequencies.
In a wide range of driving power and the thermal phonon number, the obtained entanglement degree can surpass the bound on the maximum stationary entanglement ${\rm ln}2$ from the parameter interaction.
Moreover, we have also shown that two mechanical modes
can be cooled down simultaneously by using only one driving laser. This ensures that the obtained entangled state has high purity, and can be used to implement continuous-variable teleportation with high fidelity.
This study provides a promising route to realize strong entanglement between two macroscopic systems and has potential applications in quantum information science in the future.

\begin{acknowledgements}
This work is supported by the National Key Research and Development Program of China grant 2016YFA0301200, the National Science Foundation of China (Grant Nos. 11374116, 11574104 and 11375067), and the YDW acknowledge the support from Chinese Youth 1000 Talents Program and the NSFC grants (No. 11574330 and No. 11434011).
\end{acknowledgements}


\begin{thebibliography}{99}
\bibitem{Horodecki2009} R. Horodecki, P. Horodecki, M. Horodecki, and K. Horodecki, Rev. Mod. Phys. {\bf 81}, 865 (2009).

\bibitem{Braunstein2005} S. L. Braunstein and P. v. Loock, Rev. Mod. Phys. {\bf 77}, 513 (2005).

\bibitem{Jones2012} J. A. Jones and D. Jaksch, Quantum Information, Computation and Communication (Cambridge University Press, Cambridge, U.K., 2012).

\bibitem{Giovannetti2011} V. Giovannetti, S. Lloyd, and L. Maccone, Nat. Photon. {\bf 5}, 222 (2011).

\bibitem{Reid2009} M. D. Reid, P. D. Drummond, W. P. Bowen, E. G. Cavalcanti, P. K. Lam, H. A. Bachor, U. L. Andersen, and G. Leuchs, Rev. Mod. Phys. {\bf 81}, 1727 (2009).

\bibitem{Pan2012} J.-W. Pan, Z.-B. Chen, C.-Y. Lu, H.
Weinfurter, A. Zeilinger, and M. zukowski, Rev. Mod. Phys. {\bf 84}, 777 (2012).

\bibitem{Raimond2001} J. M. Raimond, M. Brune, and S. Haroche, Rev. Mod. Phys. {\bf 73}, 565 (2001).

\bibitem{You2011} J. Q. You, and F. Nori, Nature {\bf 474}, 589 (2011).

\bibitem{Schwab2005} K. C. Schwab and M. L. Roukes, Phys. Today {\bf 58}, 36 (2005).

\bibitem{Modi2012} K. Modi, A. Brodutch, H. Cable, T. Paterek, and V. Vedral, Rev. Mod. Phys. {\bf 84}, 1655 (2012).

\bibitem{Eisert2004} J. Eisert, M. B. Plenio, S. Bose, and J. Hartley, Phys. Rev. lett. {\bf 93}, 190402 (2004).

\bibitem{Xue2007} F. Xue, Y.-X. Liu, C. P. Sun, F. Nori, Phys. Rev. B {\bf 76}, 064305 (2007).

\bibitem{Roncaglia2008} J. P. Paz and A. J. Roncaglia, Phys. Rev. lett. {\bf 100}, 220401 (2008).

\bibitem{Huang2009} S.m. Huang, and G. S. Agarwal, New J. Phys. {\bf 11}, 103044 (2009).

\bibitem{Jost2009} J. D. Jost, J. P. Home, J. M. Amini, D. Hanneke, R. Ozeri, C. Langer, J. J. Bollinger, D. Leibfried and D. J. Wineland, Nature {\bf 459}, 683 (2009).

\bibitem{Ludwig2010} M. Ludwig, K. Hammerer, and F. Marquardt, Phys. Rev. A {\bf 82}, 012333 (2010).

\bibitem{Cohen2013} G. Z. Cohen, M. DiVentra, Phys. Rev. B {\bf 87}, 014513 (2013).

\bibitem{Walter2013} S. Walter, J. C. Budich, J. Eisert, and B. Trauzettel, Phys. Rev. B {\bf 88}, 035441 (2013).

\bibitem{xinyou2013} X.-Y. L\"{u}, Z.-L. Xiang, W. Cui, J. Q. You, and F. Nori, Phys. Rev. A {\bf 88}, 012329 (2013).

\bibitem{Szorkovszky2014} A. Szorkovszky, A. A. Clerk, A. C. Doherty, and W. P. Bowen, New J. Phys. {\bf 16}, 063043 (2014).

\bibitem{Johansson2014} J. R. Johansson, N. Lambert, I. Mahboob, H. Yamaguchi, and F. Nori, Phys. Rev. B {\bf 90}, 174307 (2014).

\bibitem{Review1} M. Aspelmeyer, T. J. Kippenberg, and F. Marquardt, Rev. Mod. Phys. {\bf 86}, 1391 (2014);
M. Aspelmeyer, P. Meystre, and K. Schwab, Phys. Today {\bf 65}, 29 (2012);
P. Meystre, Ann. Phys. (Berlin) {\bf 525}, 215 (2013); F. Marquardt and S. M. Girvin, Physics {\bf 2}, 40 (2009); T. J. Kippenberg and K. J. Vahala, Science {\bf 321}, 1172 (2008).

\bibitem{Review2} H. Xiong,  L.-G. Si, X.-Y. L\"{u}, X.-X. Yang, Y. Wu, Science China Physics, Mechanics and Astronomy 58, 1-13 (2015).

\bibitem{Review3} C.-P. Sun, Y. Li, Science China Physics, Mechanics and Astronomy 58, 050300 (2015).

\bibitem{Mancini2002} S. Mancini, V. Giovannetti, D. Vitali, and P. Tombesi, Phys. Rev. lett. {\bf 88}, 120401 (2002).

\bibitem{Hartmann2008} M. J. Hartmann, and M. B. Plenio, Phys. Rev. lett. {\bf 101}, 200503 (2008).

\bibitem{Vacanti2008} G. Vacanti, M. Paternostro, G. M. Palma, and V. Vedral1, New J. Phys. {\bf 10}, 095014 (2008).

\bibitem{Zhou2011} L. Zhou, Y. Han, J. Jing, and W. Zhang, Phys. Rev. A  {\bf 83}, 052117 (2011).

\bibitem{Joshi2012} C. Joshi, J. Larson, M. Jonson, E. Andersson, and P. Ohberg, Phys. Rev. A  {\bf 85}, 033805 (2012).

\bibitem{Palomaki2013} T. A. Palomaki, J. D. Teufel, R. W. Simmonds, K. W. Lehnert, Science {\bf 342}, 710 (2013).

\bibitem{Wang2015} H. Wang, X. Gu, Y.-x. Liu, A. Miranowicz, and F. Nori, Phys. Rev. A {\bf 92}, 033806 (2015).

\bibitem{Feng2015} Q. Wang, J.-Q. Zhang, P.-C. Ma, C.-M. Yao, and M. Feng,  Phys. Rev. A {\bf 91}, 063827 (2015).

\bibitem{Sete2014} E. A. Sete and H. Eleuch, Phys. Rev. A  {\bf 89}, 013841 (2014).

\bibitem{Liao2014} J.-Q. Liao, Q.-Q. Wu, and F. Nori, Phys. Rev. A {\bf 89}, 014302 (2014).

\bibitem{Chen2014} R.-X. Chen, L.-T. Shen, Z.-B. Yang, H.-Z. Wu, S.-B. Zheng, Phys. Rev. A {\bf 89}, 023843 (2014).

\bibitem{Abdi2015} M. Abdi, M. J Hartmann, New J. Phys. {\bf 17}, 013056 (2015).

\bibitem{Girvin2011} K. B{\o}rkje, A. Nunnenkamp, and S. M. Girvin, Phys. Rev. lett. {\bf 107}, 123601 (2011).

\bibitem{Zubairy2013} W. Ge, M.Al-Amri, H. Nha, and M. S. Zubairy, Phys. Rev. A {\bf 88}, 022338 (2013).

\bibitem{Zubairy} W. Ge, M.Al-Amri, H. Nha, and M. S. Zubairy, Phys. Rev. A {\bf 88}, 052301 (2013).

\bibitem{SeteJOSAB} E. A. Sete, H. Eleuch, and C. H. Raymond Ooi, J. Opt. Soc. Am. B {\bf 31}, 2821-2828 (2014).

\bibitem{Sete2015} E. A. Sete and H. Eleuch, J. Opt. Soc. Am. B {\bf 32}, 971-982 (2015).

\bibitem{Tian2013} L. Tian, Phys. Rev. Lett. {\bf 110}, 233602 (2013).

\bibitem{Wang2013} Y.-D. Wang, A. A. Clerk, Phys. Rev. lett. {\bf 110}, 253601 (2013).

\bibitem{Tan2013} H. Tan, G. Li, P. Meystre, Phys. Rev. A {\bf 87}, 033829 (2013).

\bibitem{Woolley2014} M. J. Woolley, A. A. Clerk, Phys. Rev. A {\bf 89}, 063805 (2014).

\bibitem{Tian2008} L. Tian, M. S. Allman, R. W. Simmonds, New J. Phys. {\bf 10}, 115001 (2008).

\bibitem{Lv2015} X.-Y. L\"{u}, J.-Q. Liao, L. Tian, and F. Nori, Phys. Rev. A {\bf 91}, 013834 (2015).

\bibitem{Westra2010} H. J. R. Westra, M. Poot, H. S. J. van der Zant, W. J. Venstra, Phys. Rev. lett. {\bf 105}, 117205 (2010).

\bibitem{Westra2011} H. J. R. Westra, D. M. Karabacak, S. H. Brongersma, M. Crego-Calama, H. S. J. van der Zant, and W. J. Venstra, Phys. Rev. B {\bf 84}, 134305 (2011).

\bibitem{Okamoto2013} H. Okamoto, A. Gourgout, C.-Y. Chang, K. Onomitsu, I. Mahboob, E. Y. Chang, H. Yamaguchi, Nature Phys. {\bf 9}, 2665 (2013).

\bibitem{Liao2016} J.-Q. Liao and L. Tian, Phys. Rev. lett. {\bf 116}, 163602 (2016).

\bibitem{Kippenberg2007} I. Wilson-Rae, N. Nooshi, W. Zwerger, and T. J. Kippenberg, Phys. Rev. Lett. {\bf 99}, 093901 (2007).

\bibitem{Marquardt2007} F. Marquardt, J. P. Chen, A. A. Clerk, and S. M. Girvin, Phys. Rev. Lett. {\bf 99}, 093902 (2007).

\bibitem{Vitali2008} C. Genes, D. Vitali, P. Tombesi, S. Gigan, and M. Aspelmeyer, Phys. Rev. A {\bf 77}, 033804 (2008).

\bibitem{Vidal2002} G. Vidal and R. F. Werner, Phys. Rev. A {\bf 65}, 032314 (2002).

\bibitem{Braunstein1998} S. L. Braunstein and H. J. Kimble, Phys. Rev. Lett. {\bf 80}, 869 (1998).

\bibitem{Fiurasek2002} J. Fiurasek, Phys. Rev. A {\bf 66}, 012304 (2002).
\end{thebibliography}
\end{document}